\documentclass[prd,twocolumn,a4paper,superscriptaddress,floatfix,nofootinbib]{revtex4}
\usepackage{graphicx}
\usepackage{amssymb}
\usepackage{amsmath}
\usepackage{makecell,multirow}
\usepackage{hyperref}

\begin{document}
%My commands

\title{Detection and Characterization of Quasi-Periodic Oscillations in Seyfert Galaxy NGC 4151}

\author{Yan Yongkang}
\affiliation{College of Mathematics and Physics, China Three Gorges University, Yichang 443002, China}

\author{Zhang Peng}
\email[Electronic address: ]{zhangpeng@ctgu.edu.cn}
\affiliation{College of Mathematics and Physics, China Three Gorges University, Yichang 443002, China}
\affiliation{Center for Astronomy and Space Sciences, China Three Gorges University, Yichang 443002, China}

\author{Lu Zhou}
\affiliation{College of Mathematics and Physics, China Three Gorges University, Yichang 443002, China}
\affiliation{Center for Astronomy and Space Sciences, China Three Gorges University, Yichang 443002, China}

\author{Bao Tong}
\affiliation{INAF – Osservatorio Astronomico di Brera, Via E. Bianchi 46, 23807 Merate (LC), Italy}

\author{Liu Dejian}
\affiliation{College of Mathematics and Physics, China Three Gorges University, Yichang 443002, China}
\affiliation{Center for Astronomy and Space Sciences, China Three Gorges University, Yichang 443002, China}

\author{Liu Gaochao}
\affiliation{College of Mathematics and Physics, China Three Gorges University, Yichang 443002, China}
\affiliation{Center for Astronomy and Space Sciences, China Three Gorges University, Yichang 443002, China}

\author{Liu Qingzhong}
\affiliation{Purple Mountain Observatory, Chinese Academy of Sciences, Nanjing 210008, China}

\author{Yan Jingzhi}
\affiliation{Purple Mountain Observatory, Chinese Academy of Sciences, Nanjing 210008, China}

\author{Zeng Xiangyun \thanks{xyzeng2018@ctgu.edu.cn}}
\email[Electronic address: ]{xyzeng2018@ctgu.edu.cn}
\affiliation{College of Mathematics and Physics, China Three Gorges University, Yichang 443002, China}
\affiliation{Center for Astronomy and Space Sciences, China Three Gorges University, Yichang 443002, China}

\begin{abstract}
This study aims to detect and characterize quasi-periodic oscillations (QPOs) signals in X-ray observations of NGC 4151. We employed the Weighted Wavelet Z-transform (WWZ) and Lomb-Scargle periodogram (LSP) methods for our analysis. QPO signals with frequencies of 5.91 $\times 10^{-4}$ Hz and 5.68 $\times 10^{-4}$ Hz were detected in observations conducted by Chandra (ObsID 7830) in 2007 and XMM-Newton (ObsID 0761670301) in 2015, with confidence levels of 3.7 $\sigma$ and 3.3 $\sigma$, respectively. These signals are the first to be independently observed by two different telescopes over an eight-year period with closely matched frequencies. Most notably, the combined confidence level of the QPO signals from these two independent observations reaches an exceptional 5.2 $\sigma$, which is rare in astrophysical research and significantly strengthens our conviction in the authenticity of these signals. A detailed analysis of the observational data suggests that these QPO signals may be correlated with the properties of the central supermassive black hole. Additionally, spectral analysis of the observational data revealed no significant spectral differences between the QPO and non-QPO segments. These findings provide new insights into the X-ray variability mechanisms of the central black hole in NGC 4151 and offer a novel perspective for black hole mass estimation.

\end{abstract} 
\maketitle

\section{Introduction}
\label{sec:intro}

Active galactic nuclei (AGN) are extragalactic objects in which supermassive black holes (BH) release intense energy during the accretion of matter. The anisotropic structures within these galactic nuclei endow them with a variety of observational characteristics across the entire electromagnetic spectrum \cite{Swain2023}. The X-ray spectra of AGN primarily originate from regions close to the black hole within the accretion disk \cite{Ghisellini1994}, and their radiative variability reveals the dynamic changes in the accretion process and the shielding effects of the environment \cite{Lachowicz2005}. Quasi-Periodic Oscillations (QPO) are a special phenomenon in X-ray and $\gamma$-ray emitting sources, manifesting as periodic fluctuations in X-ray light curves, with frequencies that are not constant, commonly observed in neutron star or black hole binaries in the Milky Way and nearby galaxies \cite{Gierlinski2008, Pan2016}, but rare in AGN \cite{Zhang2020}. Many early QPO detections were discredited due to insufficient modeling of potential broadband noise \cite{Alston2015}. 

NGC 4151, a quintessential Seyfert Type 1 galaxy, is known as the "Eye of Sauron" due to its unique and striking appearance. Its significant brightness and variability across multiple wavelengths make it one of the most extensively studied objects in the electromagnetic spectrum, providing a crucial testing ground for numerous theoretical models. The host galaxy of NGC 4151 is classified as an intermediate barred spiral (type SABab), located in the constellation Canes Venatici. There is considerable debate regarding the distance and central black hole mass of NGC 4151, but the widely accepted distance is $19.0^{+2.4}_{-2.6}$ Mpc \cite{Sebastian2014}, and its redshift is determined to be $z=0.003152$ \cite{Koss2022}. The central black hole mass estimates range from $2.5 \times 10^{6} M_{\odot}$ to $5.6 \times 10^{7} M_{\odot}$, for a comprehensive discussion on determining black hole mass using various indirect methods, see refs. \cite{Williams2023}. Photometry observations of NGC 4151 date back to 1910, with initial suggestions of a 13.7 year periodicity in its long-term variability analyzed by refs. \cite{Pacholczyk1983} through power spectral analysis of early optical data. Subsequent analysis refined this estimate to 14.08 $\pm$ 0.8 years. However, studies on short-term variability in NGC 4151 are relatively scarce. Understanding this variability is crucial for advancing our knowledge of the dynamic processes occurring in the vicinity of the central black hole and their impact on the broader electromagnetic spectrum. The complex interplay between the black hole's mass, accretion disk, and observed X-ray emissions remains a topic of significant astrophysical interest.

The study employs the Weighted Wavelet Z-transform (WWZ) and Lomb-Scargle periodogram (LSP) to detect and characterize QPO signals in the X-ray observations of NGC 4151. The findings provide insights into the ongoing discussion regarding the transient nature of QPOs and their potential correlation with the properties of the central black hole. The structure of this paper is as follows: Section 2 details the observations and data processing methods, the main results are presented in Section 3, and Section 4 provides the summary and discussion.

\section{Observations and data analysis}
\label{sec:Observations}

\begin{figure*}[t]
        \centering
        \includegraphics[scale=0.4]{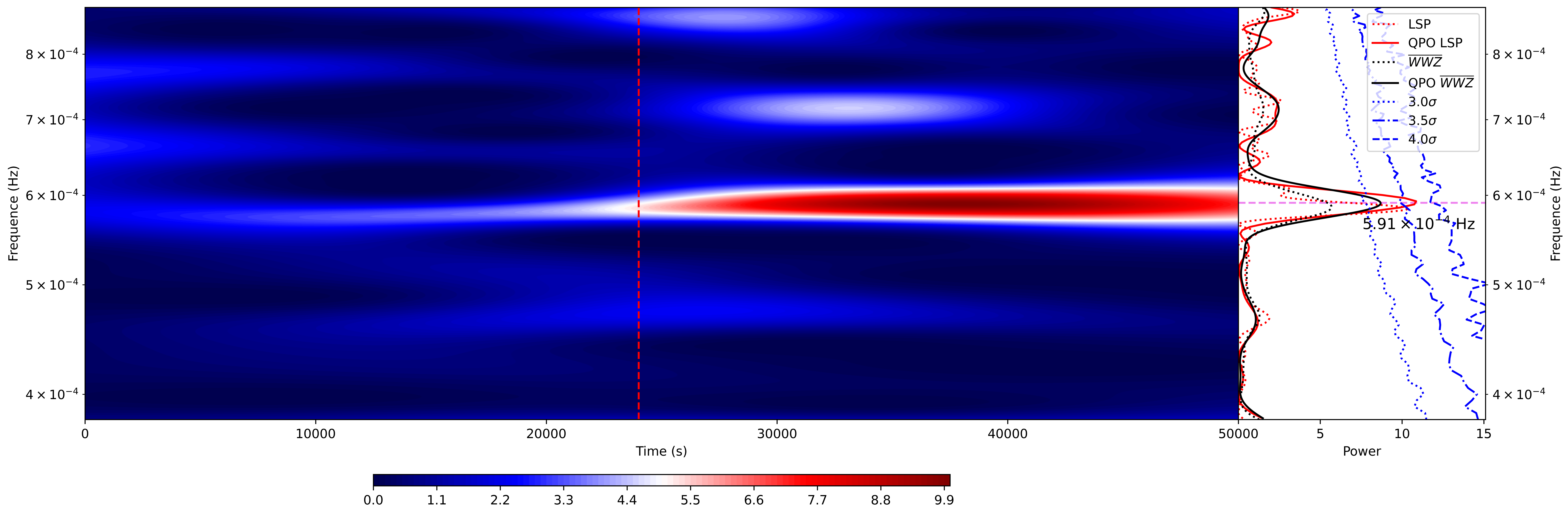}
        \caption{WWZ and LSP analysis results for 7830 observed by the ACIS instrument aboard Chandra in 2007. The left panel illustrates the variation of WWZ power with time and frequency, where the power levels are color-coded as indicated by the color bar below. The right panel presents the average power for both WWZ (black) and LSP (red). Dashed lines correspond to results averaged over the entire observation period, while solid lines represent the results from the QPO segment, demarcated by the red dashed line (at 24,050 s). Blue lines with different styles represent confidence levels of 3.0, 3.5, and 4.0\ $\sigma$.
        }
        \label{7830}
\end{figure*}

\begin{figure*}
        \centering
        \includegraphics[scale=0.4]{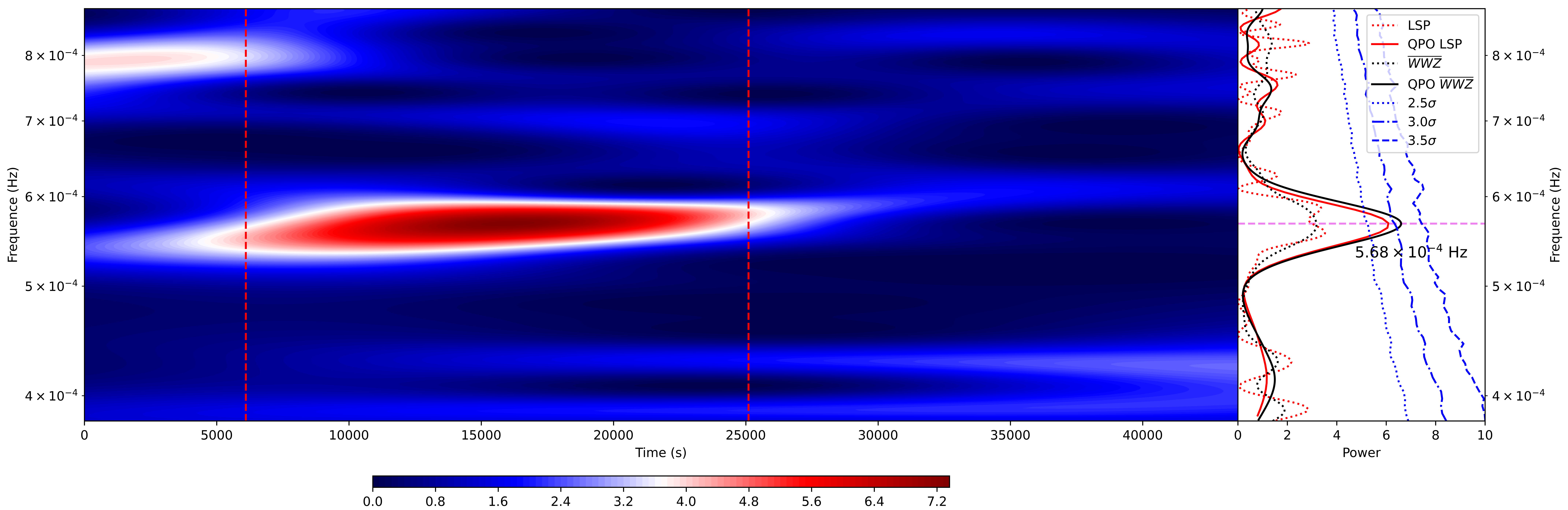}
        \caption{WWZ and LSP analysis results from the EPIC camera data of XMM-Newton 0301 observed in 2015. The red dashed line indicates the segmentation, with the interval from 6,150 s to 25,150 s identified as the QPO segment. Blue curves with varying styles denote confidence levels of 2.5, 3.0, and 3.5\ $\sigma$. Other features are consistent with those described in Figure \ref{7830}.
        }
        \label{0761670301}
\end{figure*}

XMM-Newton X-ray observatory, launched in 1999, is equipped with two sets of X-ray detectors, including three European Photon Imaging Cameras (EPIC: PN, MOS1, and MOS2; \cite{Strder2001, Turner2001}) and two Reflection Grating Spectrometers (2RGS; \cite{Herder2001}). Chandra X-ray Observatory \cite{Weisskopf2002}, launched by NASA on July 23, 1999, is a high-energy X-ray observatory designed to study the hottest and most active objects in the universe, particularly X-ray emitting sources such as black holes, neutron stars, AGNs, and supernova remnants \cite{Weisskopf2012}. Chandra is capable of detecting X-rays in the 0.1 keV to 10.0 keV range with exceptional sensitivity, allowing it to observe faint X-ray emissions from distant galaxy clusters. Using its high-resolution camera (HRC) and advanced CCD imaging spectrometer (ACIS) \cite{Garmire2003}, Chandra enables precise X-ray spectral analysis. 

XMM-Newton has performed 32 observations of NGC 4151, accumulating a total observing time of 1,276 ks. The main data analyzed in this paper were obtained from a 47 ks observation conducted on November 16, 2015, with ObsID 0761670301 (hereafter referred to as 0301). The data processing method for XMM-Newton has been described in detail in refs. \cite{Yan2024}, and thus is not repeated here.

Chandra has conducted 18 observations of NGC 4151, with a total observing time of 824 ks, including 14 long observations ($\textgreater$ 30 ks). The primary data used in this study come from a 49.34 ks ACIS-S observation conducted on July 21, 2007, with ObsID 7830 (hereafter referred to as 7830). The data were processed in accordance with the standard Chandra ACIS data preparation procedures, utilizing the Chandra X-ray Center CIAO v4.16.0\footnote{\url{https://cxc.harvard.edu/ciao/}} software and the CALDB v4.11.0\footnote{\url{https://cxc.harvard.edu/caldb/}} calibration files. Initially, all the ObsIDs for NGC 4151 were retrieved using the \textit{find\_chandra\_obsid} tool, followed by the download of observational data using the \textit{download\_chandra\_obsid} tool. Preprocessing was then performed using \textit{chandra\_repro}, generating level 2 event files. A circular region of interest (ROI) was selected, centered at right ascension (R.A.) $=$12$^h$10$^m$32$^s$.61, declination (Dec.) $=$ 39$^{\circ}$24$^{\prime}$20$^{\prime\prime}$7, with the radius determined according to the guidelines provided in Chandra Proposers’ Observatory Guide, v26.0\footnote{\url{https://cxc.harvard.edu/proposer/POG/html/index.html}}. To minimize errors in the data, the energy range for event extraction was restricted to 0.2 -- 10.0 keV, based on the total response efficiency of Chandra in relation to incident photon energy \cite{Bao2024}. The time bins were set to 100 s. Scientific data were processed using the \textit{dmextract} tool to generate the light curve data. To eliminate errors caused by background photons, a source-free circular ROI from the observation, which had no X-ray sources, was selected and used as the background data for subtraction. Subsequent time series analysis was based on the light curve data obtained in this manner.

Finally, the source and background pulse invariant (PI) spectra, as well as the associated auxiliary response file (ARF) and redistribution matrix file (RMF), were generated using the same ROI data and the \textit{specextract} tool. To examine the spectral differences between regions with and without QPO, the \textit{dmcopy} tool was used to select the required specific time intervals for event extraction. Additionally, the \textit{grppha} tool was employed to merge data points, thereby reducing overall errors and avoiding the influence of high-error data points, preparing the data for subsequent spectral analysis.

\section{Result}
\subsection{Timing results}
\label{subsec:searchqpo}

Wavelet transforms have demonstrated significant advantages as a method for periodicity analysis in time series, particularly for detecting the time evolution of parameters (period, amplitude, phase) that describe periodic and quasi-periodic signals. By considering the wavelet transform as a projection, its statistical behavior was derived, and favorable rescaling transformations were designed. By treating it as a weighted projection to form WWZ, enhancing its capability to detect periodic and quasi-periodic signals. In this study, Morlet function were used for the mother function of WWZ.

To further investigate and quantify periodicity, LSP \cite{Lomb1976, Scargle1982}, a method that differs from WWZ in both principle and application, was additionally utilized. This widely-used technique for power spectral analysis of unevenly sampled time series data has the advantage of being able to analyze periodicity in irregular time series. It uses a $\chi^2$ statistic to fit sine waves across the entire data set, thereby reducing the impact of irregular sampling. The combination of LSP and WWZ can provide a more comprehensive method for detecting quasi periodic signals. 

WWZ analysis was performed on the obtained light curves, resulting in two-dimensional color maps that depict the variation of the power spectrum over time and frequency. These maps provide a clear and intuitive view of how the power spectrum evolves, enabling the identification of the quasi-periodic locations and their corresponding intensities. Figure \ref{7830} displays the processing results for the 7830. A peak is detected at (5.91 $\pm$ 0.21) $\ \times\ $10$^{-4}$\ Hz (1,692 $\pm$ 59 s, $\sim$ 0.47 hours, with the error representing the full width at half maximum (FWHM)), observed in both the WWZ and LSP power spectra. The dashed lines represent the WWZ and LSP results over the entire time period, where the WWZ appears relatively lower due to its averaging of the power over the full observation time. Based on the periodicity observed in the two-dimensional map, the data were further segmented at 24,050 s. The solid lines indicate the results after this segmentation for the QPO region, where a significant increase in the WWZ value is clearly observed, whereas the LSP value shows little enhancement. This is because the WWZ is more sensitive to localized time periods where the signal is present, whereas methods like LSP, which consider the entire time span, are less sensitive to such localized variations.

For 0301, the results are shown in Figure \ref{0761670301}. A peak is observed at (5.68 $\pm$ 0.23) $\ \times\ $10$^{-4}$\ Hz (1760 $\pm$ 70 s, $\sim$ 0.49 hours, with the error representing FWHM). By examining the two-dimensional map, it is evident that the QPO signal is concentrated between 6,150 s and 25,150 s (the middle region surrounded by the red line). From the light curve, it can also be seen that the portion containing the QPO exhibits clear periodicity, whereas the rest of the data is more chaotic. Further segmentation of the light curve, similarly shown by the solid lines, indicates the results after processing the segmented QPO section.

To evaluate the reliability of the detected signal and to estimate its confidence level more accurately, the probability of the signal being generated by stochastic noise was further assessed. Using the observed power spectral density (PSD) of the light curve and its probability density function, we employed the algorithm provided by Emmanoulopoulos \citep{Emmanoulopoulos2013} to generate $10^6$ simulated stochastic light curves. To determine the optimal fit for the PSD, the PSD of the observed light curve was modeled using a bending power-law with a constant component, and the fitting procedure was performed using the Minuit $\chi^2$ minimization technique. The adopted function is expressed as $P(f) = A f^{-1} \left[ 1 + \left( \frac{f}{f_{bend}} \right)^{\alpha - 1} \right]^{-1} + C$ \citep{Vaughan2012},where $A$ , $ \alpha$(divided into two values: $\alpha_{\text{low}}$ for the low-frequency range and $\alpha_{\text{high}}$ for the high-frequency range) , $ f_{bend}$ and $C$ denote the normalization, the spectral index above the bending frequency, the bending frequency, and the Poisson noise level, respectively. The fitting results are as follows: for 7830, $A=9.79 \times 10^{-8}$, $\alpha_{\text{low}}=1.60$, $\alpha_{\text{high}}=1.03$, $f_{\text{bend}}=0.0198$, $C=2.73$; for 0301, $A=3.06 \times 10^{-10}$, $\alpha_{\text{low}}=2.26$, $\alpha_{\text{high}}=2.35$, $f_{\text{bend}}=0.0759$, $C=0.168$. Based on this methodology, the influence of stochastic components such as red noise and white noise was accounted for. Following the procedure outlined by \cite{Emmanoulopoulos2013}, artificial light curves with identical binning and duration were generated, and the corresponding confidence contours were computed to evaluate the observed signal's significance.

%\begin{table}
%    \renewcommand{\arraystretch}{1.1} % 调整行间距
%    \centering
%    \caption{Best-fit Parameters for the Power Spectral Density Model}
%    \begin{tabular}{cccccc}
%        \hline
%        ObsID & $A$ & $\alpha_{\text{low}}$ & $\alpha_{\text{high}}$ & $f_{\text{bend}}$ & $C$ \\
%         \hline
%        7830 & $9.79 \times 10^{-8}$ & 1.60 & 1.03 & 0.0198 & 2.73 \\
%        0761670301 & $3.06 \times 10^{-10}$ & 2.26 & 2.35 & 0.0759 & 0.168 \\
%         \hline
%    \end{tabular}
%    \label{psd}
%\end{table}

In the right panels of Figures \ref{7830} and \ref{0761670301}, the confidence curves are plotted in blue with different line styles, corresponding to 3.0, 3.5, and 4.0 $\sigma$ confidence levels. A rough estimate of the confidence can be made by visually inspecting the figures. After more precise calculations, it was found that the QPO signal in 7830 had a confidence level of 3.7 $\sigma$, while the 0301 showed a confidence level of 3.3 $\sigma$. Both values exceed the threshold of 3 $\sigma$, which is typically used to identify a genuine signal. Therefore, it is concluded that QPO signals with nearly identical frequencies were detected in both independent observations.

The two QPO signals were observed in Chandra (7830) in 2007 and XMM-Newton (0301) in 2015, marking the first time that AGN QPO signals with such close frequencies have been detected by two different telescopes, with an observational gap of eight years. It is observed that the central frequency of the signal in Chandra is slightly higher than that in XMM-Newton, although the difference is still within the error range, implying that the deviation is not statistically significant. Similar to the observations reported in refs. \cite{Song2020}, the analysis also reveals that QPO signals with nearly identical frequencies were detected in two independent observations, spaced 8 years apart, using two different telescopes. The combined blind-chance probability $P_{false}$ of these two QPO signals arising purely from random noise is 7.57 $\times 10^{-8}$, corresponding to a 99.999992$\%$ confidence level, or approximately 5.2 $\sigma$. In the extreme case of complete non-independence, the combined confidence would be the higher of the two values, about 3.7 $\sigma$, which still provides sufficient statistical significance to confirm the existence of the QPO signal. Furthermore, a similar phenomenon has been observed in RE J1034+396, where a QPO signal with a nearly identical period was detected years later using the same telescope (\emph{XMM-Newton}), with a high level of confidence \citep{Jin2020}.

\subsection{spectral analysis}

To further investigate the spectral properties of NGC 4151, spectral analysis of 7830 was performed using XSPEC (Version 12.14.0n, \cite{Arnaud1996}). Considering the total response efficiency of Chandra in relation to the energy of incident photons, and to minimize potential data errors, the analysis was restricted to the 1.0 keV -- 10.0 keV energy range. In this model, \textit{zpowerlaw} is a redshift-corrected variant of the simple power law used to represent the continuum spectrum. It incorporates the redshift factor, with $z=0.003152$ \cite{Koss2022}. \textit{TBabs} is the Tuebingen-Boulder ISM absorption model, which accounts for Galactic absorption in NGC 4151. Based on refs. \cite{Couto2016},The $N_H$ parameter was set at $1.98 \times 10^{20}\, \text{cm}^{-2}$. The model provides a reasonable fit for the hard X-ray segment, with an overall goodness of fit of $\chi^{2}/\nu = 1.12$.

\begin{figure}
        \centering
        \includegraphics[scale=0.35]{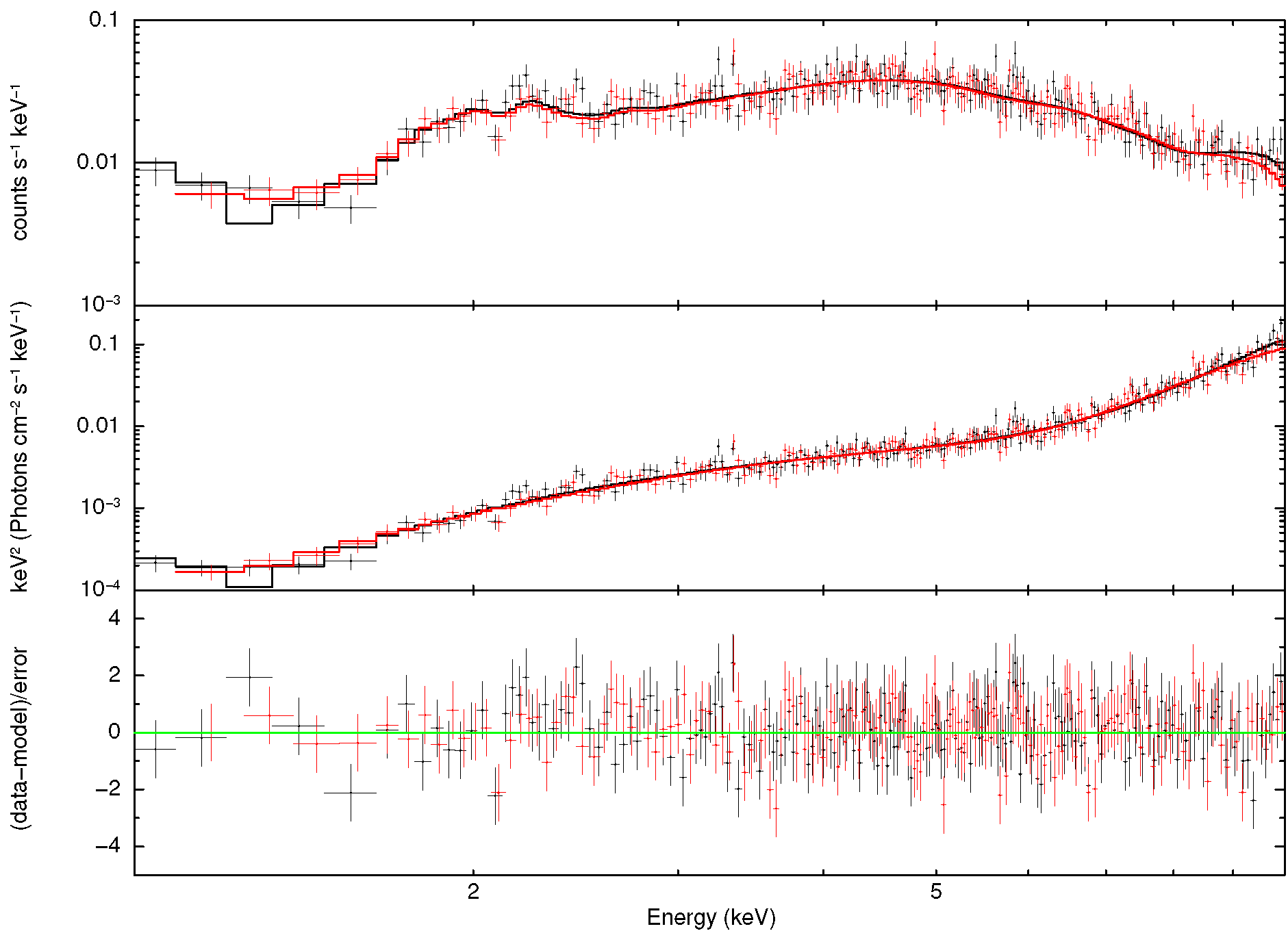}
        \caption{Spectrum and fitting results of Chandra 7830. The black data points represent the QPO segment, while the red points correspond to the non-QPO segment. The upper and middle panel shows the ld and eeuf plot, respectively. The lower panel presents the residuals of the best-fit model, calculated as [data–model] / error.
        }
        \label{spectrum}
\end{figure}

The energy spectrum of NGC 4151 frequently exhibits the Fe $K\alpha$ emission line at 6.315 keV, as extensively studied in refs. \cite{Kraemer2020, Collaboration2024, Wang2010}. However, it is noteworthy that this emission line is absent in the 7830 analyzed in this study. Instead, as shown in the spectral plot (Figure \ref{spectrum}), an absorption feature around 6 keV is observed, which is consistent with the findings reported in refs. \cite{Couto2016}. To account for this, the \textit{gabs} model was incorporated into the fitting process. The results demonstrate that this model effectively describes the hard-energy band, yielding a well-distributed residual. However, significant deviations remain in the soft-energy range between 1.0 and 2.0 keV. To address this, the \textit{edge} model was further introduced to account for edge absorption. The addition of this model improved the fit considerably. Consequently, the final model adopted for this analysis is defined as \textit{TBabs $\times$ zpowerlaw $\times$ gabs $\times$ edge}.

Further investigation of the spectral differences between the QPO and non-QPO segments was conducted by extracting the spectra for these intervals using Good Time Intervals (GTIs). Both datasets were fitted using the same model and parameters described earlier. The fitting results indicate no significant differences between the two spectra; they exhibit similar X-ray luminosities and spectral shapes. The best-fit spectral models are shown in the middle panel of Figure \ref{spectrum}, where the black curve represents the QPO segment and the red curve corresponds to the non-QPO segment. The top panel displays the logarithmic data (ld), the middle panel shows the unfolded energy spectrum (eeuf), and the bottom panel presents the residuals, which are evenly distributed, indicating a good fit between the model and the observed data.

\begin{table}
    \renewcommand{\arraystretch}{1.3} % 调整行间距
    \centering
    \caption{Best-fitting spectral parameters of Chandra 7830}
    \begin{tabular}{c@{\hskip 10pt}l@{\hskip 10pt}c@{\hskip 10pt}c}
        \hline
        \textbf{Component} & \textbf{Parameter}       & \textbf{QPO}                        & \textbf{Non-QPO} \\ 
        \hline
        \multirow{3}{*}{gabs} 
             & $E_{line}$/keV			             & $6.08_{-0.23}^{+0.20}$              & $5.96_{-0.18}^{+0.22}$ \\ 
             & $\sigma_{line}$/keV	 			            & $2.46_{-0.50}^{+1.20}$              & $1.81_{-0.33}^{+0.40}$ \\ 
             & $d_{line}$/keV	 		           & $10.02_{-3.98}^{+2.90}$            & $4.78_{-1.49}^{+3.61}$ \\ 
        \hline
        \multirow{2}{*}{zpowerlw} 
             & $\Gamma$                 & $-0.86_{-0.21}^{+0.20}$             & $-0.80_{-0.17}^{+0.18}$ \\ 
             & Norm ($\times 10^{-4}$)  & $2.61_{-0.96}^{+2.66}$              & $1.57_{-0.44}^{+0.87}$ \\ 
        \hline
        \multirow{2}{*}{edge} 
             & $E_{edge}$/keV				             & $1.17_{-0.05}^{+0.09}$              & $1.17_{-1.17}^{+0.16}$ \\ 
             & $\tau_{max}$                   & $1.74_{-0.52}^{+0.63}$              & $0.76_{-0.76}^{+0.65}$ \\ 
        \hline
        \multirow{1}{*}{Reduced} 
             & $\chi^2/\nu$             & $1.04$                              & $1.03$ \\ 
        \hline
    \end{tabular}
\begin{center}
\footnotesize{\textbf{Note.} The errors represent the $3.0\ \sigma$ confidence limits for a given parameter.}
\end{center}
    \label{model}
\end{table}

It is worth noting that in this study, the power-law model used to fit the data adopts the photon index ($\Gamma$) convention rather than the spectral index($\alpha$). The photon index typically refers to the negative slope of the spectrum, expressed as $F(E) \propto E^{-\Gamma}$. In some literature (e.g., refs. \cite{Zdziarski1996}), particularly in models involving thermal emission, the spectral index is also used to describe the slope of the spectrum. The relationship between the two indices is given by $\Gamma = \alpha + 1$. Therefore, in certain cases, the photon index and spectral index can be interchanged (see refs. \cite{Ishibashi2010}). In this work, the photon index ($\Gamma$) convention, which is more widely adopted, was chosen as the spectral fitting parameter to follow standard AGN X-ray spectral analysis practices and to ensure consistency with other related studies. The best-fit parameters and their associated uncertainties are listed in Table \ref{model}. Within the margin of error, the two spectra can be considered statistically identical. Furthermore, when the fit is applied to the entire observation period, the results remain consistent. It is worth noting that, compared to other observations of NGC 4151, such as those reported by refs. \cite{Wang2011c, Zoghbi2019, Gianolli2024}, the spectral data from this particular observation appear relatively simple, with fewer complex components.

\section{Summary and discussion}
\label{Results}

In this study, we processed and analyzed all available observational data of NGC 4151 obtained through the \emph{Chandra} and \emph{XMM-Newton} X-ray telescopes. The LSP method was employed to construct time-averaged power spectra, while the WWZ technique was used to generate two-dimensional color maps, aiding in the search for QPO signals. During two separate observations (7830 and 0301) QPO signals with frequencies of $5.91 \times 10^{-4}\ \mathrm{Hz}$ and $5.68 \times 10^{-4}\ \mathrm{Hz}$ were detected, with confidence levels of 3.7 $\sigma$ and 3.3 $\sigma$, respectively. When these detections are considered independent events, the combined significance of the QPO signal increases to 5.2 $\sigma$, further strengthening the evidence for the presence of QPOs in this source. Additionally, spectral fitting results indicate that the energy spectrum of this source can be well-modeled using a simple absorbed power-law model. Notably, no iron line features were observed in the spectra.

Furthermore, considering the extensive observational dataset from \emph{Chandra} and \emph{XMM-Newton}, which includes a total of 50 observations spanning a cumulative exposure time of 2,100 ks, it is notable that the QPO signal was not significantly detected in the remaining 48 observations. The observation intervals containing QPO signals thus account for only approximately $\frac{1}{44}$ of the total dataset. Specifically for \emph{Chandra}, which includes 18 individual observations, the probability of obtaining a false-positive QPO detection increases by approximately 18 times compared to a single observation segment. Under these conditions, the confidence level of the QPO signal detected in 7830 would be reduced to 2.9 $\sigma$, and similarly, the QPO signal in 0301 would decrease to 2.2 $\sigma$. However, even under these conservative assumptions and using a blind search algorithm without any prior information, the combined significance of the two signals remains as high as 4.0 $\sigma$.

When a stricter algorithm is applied—assuming that the QPO frequency is entirely unknown—the confidence level is still maintained at 3.2 $\sigma$, further substantiating the authenticity of the detected signals. Detailed results are presented in Table \ref{sigma}, where the column labeled $P_{\text{false}}$ represents the probability of the signal arising from random noise, and the column labeled $\sigma$ denotes the confidence level expressed in terms of standard deviations. It is worth noting that in all calculations throughout this work, raw data were used, and final results were rounded to two significant figures only during the output stage. Consequently, minor discrepancies may arise when recalculating from the tabulated data. 

\begin{table}   
    \renewcommand{\arraystretch}{1.2}  
    \centering  
    \caption{7830, 0301, and their combined confidence and false probability}  
    \begin{tabular}{cccc@{\hskip 8pt}cc@{\hskip 8pt}cc}  
        \hline  
        \multirow{2}{*}{ } & & \multicolumn{2}{c}{\textbf{Original}} & \multicolumn{2}{c}{\textbf{Blind}} & \multicolumn{2}{c}{\textbf{Strict}} \\  
        \cline{3-8}  
        & & \textbf{CL} & \textbf{$P_{false}$} & \textbf{CL} & \textbf{$P_{false}$} & \textbf{CL} & \textbf{$P_{false}$}  \\  
        \hline  
         7830  & & 3.7& $1.1\times10^{-4} $& 2.9& $1.9\times10^{-3} $& 2.4& $8.7\times10^{-3} $\\  
        0301 & & 3.3& $4.8\times10^{-4} $& 2.2& $1.5\times10^{-2} $& 1.5& $6.9 \times 10^{-2} $\\  
        Joint & & 5.2& $7.6\times10^{-8} $& 4.0& $3.2\times10^{-5} $& 3.2& $5.5\times10^{-4}$\\  
        \hline  
    \end{tabular}  
    \begin{center}
\footnotesize{\textbf{Note.} "CL" represents the confidence level ($\sigma$). "Original" represents the result of confidence calculation only for these two observed light curves. "Blind" represents the calculation result of blind search considering multiple observations by two telescopes, while "Strict" represents the result of blind search assuming that the QPO frequency is completely unknown.}
\end{center}
    \label{sigma}  
\end{table}

\begin{figure}
        \centering
        \includegraphics[scale=0.45]{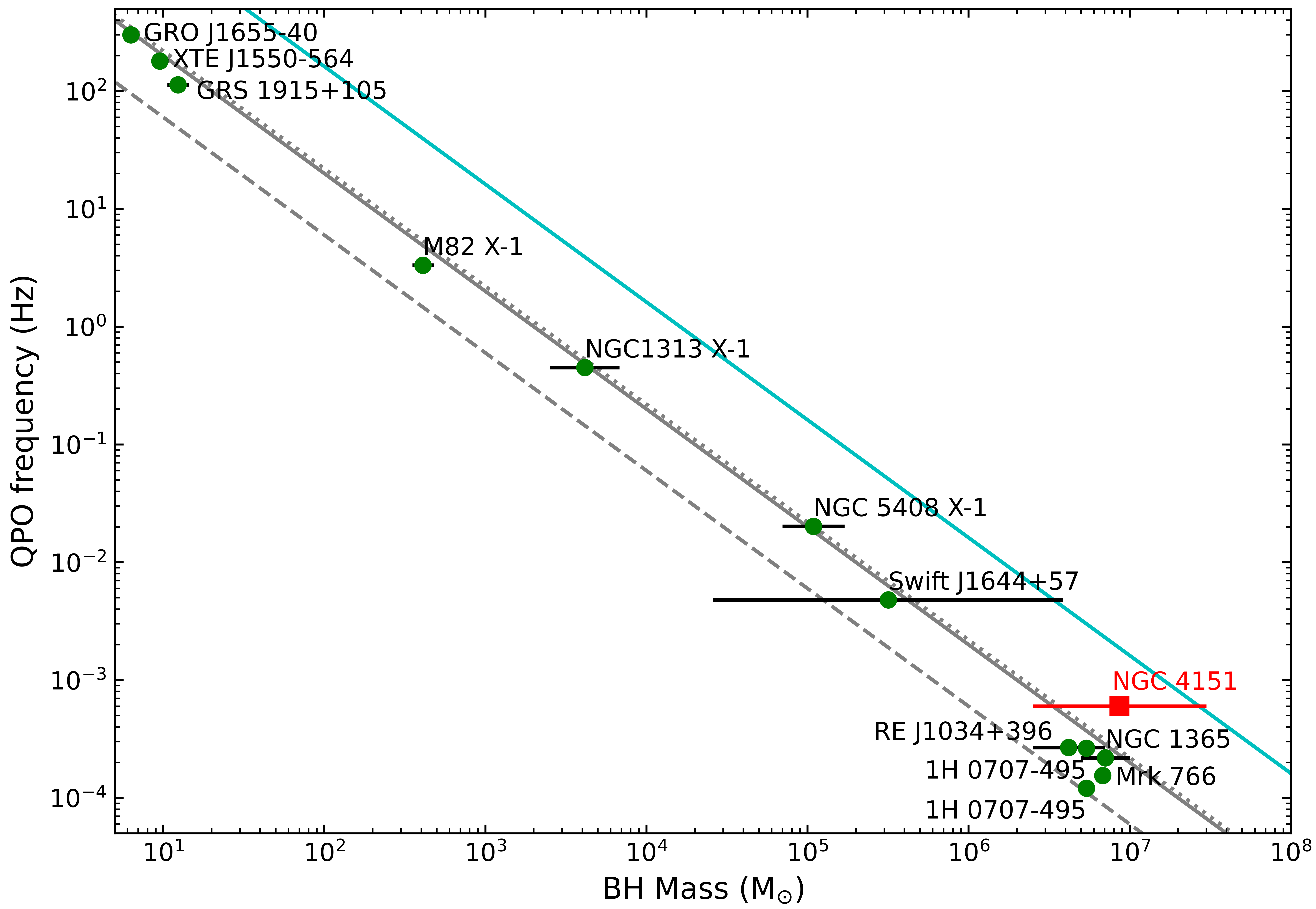}
        \caption{Correlation between black hole mass and QPO frequency. The three gray lines represent the empirical relationships proposed by refs. \cite{Remillard2006, Kluzniak2002}. The blue line reflects the maximum orbital frequency at the ISCO, as calculated by refs. \cite{Goluchova2019}, suggesting that QPO signals with frequencies exceeding the ISCO limit cannot occur in the region above and to the right of this boundary. Green circles indicate QPO signals reported in previous studies, while the newly detected QPO signal in NGC 4151 is marked with a red square. The horizontal black line denotes the estimated range of black hole mass.
        }
        \label{f-m}
\end{figure}

Unlike in X-ray binary, QPO is very difficult to detect in AGN \cite{Bao2022}. To date, QPOs that have been discovered and widely accepted in AGN include those in 1H 0707$-$495 \cite{Pan2016, Zhang2018}, NGC 1365 \cite{Yan2024}, RE J1034$+$396 \cite{Gierlinski2008}, Mrk 766 \cite{Zhang2017}, and ESO 113$-$G010 \cite{Zhang2020}. NGC 4151, like these sources, have hourly QPO periods, and QPO signals also have situations where it appears and disappears, which is very similar to X-ray binary. At present, the origin of QPO is still a controversial topic. However, existing research typically suggests that QPO signals may be generated by several mechanisms, such as pulsating accretion near the Eddington limit, instability of the inner accretion disk, X-ray hotspots around BH, and disk oscillations and precession (see refs. \cite{Syunyaev1973, Bardeen1975, Guilbert1983, Mukhopadhyay2003, Li2003, Remillard2006, Gangopadhyay2012}). Detailed explanatory models include the resonance model \cite{Abramowicz2001}, the relativistic precession model \cite{Stella1999,Zhangpf2017a,Zhangpf2017b,Zhangpf2022}, the acoustic oscillation model \cite{Rezzolla2003}, the accretion jet instability model \cite{Tagger1999}, and the Dicco seismic model assuming thin disk oscillation \cite{Wagoner1999}.

Furthermore, numerous studies have demonstrated a correlation between QPO frequency and black hole mass, providing valuable insights into the origin mechanisms of QPO phenomena. The foundational framework of these relationships and associated trend lines was initially proposed by refs. \cite{Abramowicz2004} and \cite{Torok2005}, and later refined and expanded upon by refs. \cite{Zhou2010, Zhou2015, Goluchova2019}, among others. Building upon previous studies and incorporating the new findings presented in this work, the $f_{QPO} - M_{BH}$ correlation shown in Fig. \ref{f-m} has been constructed. The plotted data were compiled from a variety of sources, including but not limited to: GRS 1915$+$105 \cite{Belloni2001}, NGC 1313 X-1 \cite{Pasham2015}, 1H 0707$-$495 \cite{Pan2016, Zhang2018}, NGC 1365 \cite{Yan2024}, RE J1034$+$396 \cite{Gierlinski2008}, and Mrk 766 \cite{Zhang2017}. As shown in the figure, this correlation holds consistently across a wide range of black hole masses, from stellar-mass black holes to supermassive black holes. For the central black hole mass of NGC 4151, this study adopts a range of estimates derived from various models used in previous works. The minimum mass estimate of $2.5 \times 10^{6} M_{\odot}$ \cite{Roberts2021} and the maximum estimate of $5.6 \times 10^{7} M_{\odot}$ \cite{Onken2014} are selected, thereby encompassing a broad range of possibilities. As a result, the overall uncertainty is expected to be relatively large.

At the same time, the newly discovered QPO signal provides an opportunity to estimate the mass of the central black hole in NGC 4151. According to refs. \cite{Goluchova2019}, the maximum QPO frequency associated with the innermost stable circular orbit (ISCO) of a rotating black hole is given by $f_{ISCO} = 16.2 \, \text{kHz} \times \frac{M_{\odot}}{M_{BH}}$. Using this relationship, the observed QPO frequency can be used to infer the upper limit for the black hole mass. Substituting the measured $f_{QPO}$ into this formula, the maximum mass of the central black hole in NGC 4151 is estimated to be $M_{max} = 2.8 \times 10^7 \, M_{\odot}$. In addition, assuming the detected QPO corresponds to a HFQPO, the black hole mass is likely to be around $3.5 \times 10^6 \, M_{\odot}$. This estimate is consistent with the mass range for NGC 4151 derived by refs. \cite{Williams2023}, who synthesized results from prior studies to provide a comprehensive assessment of the black hole's mass. The findings of this study contribute to refining the estimated mass of NGC 4151 central black hole and offer a novel perspective for black hole mass estimation.

\acknowledgments
This work is supported by the National Natural Science Foundation of China under grants 12403052, 12373030, 12203029, 12233002, U2031205, 2021YFA0718500 and 2023AFB577. The data used in this paper were obtained from XMM-Newton and Chandra, and we express our gratitude.

\bibliography{myBiblio}
\end{document}